\documentclass[onecolumn]{aa}

\usepackage{graphicx}
\usepackage{txfonts}

\begin{document}

   \title{Uncovering Erosion Effects on Magnetic Flux Rope Twist}

\author{Sanchita Pal
\and Emilia Kilpua
\and Simon Good
\and Jens Pomoell
\and Daniel J. Price}

\institute{Department of Physics, University of Helsinki, P.O. Box 64, FI-00014 Helsinki, Finland \\
email: sanchita.pal@helsinki.fi}

\date{}

  \abstract
  {Magnetic clouds (MCs) are transient structures containing large-scale magnetic flux ropes from solar eruptions. The twist of magnetic field lines around the rope axis reveals information about flux rope formation processes and geoeffectivity. During propagation, MC flux ropes may erode via reconnection with the ambient solar wind. Any erosion reduces the magnetic flux and helicity of the ropes, and changes their cross-sectional twist profiles.}
  {This study relates twist profiles in MC flux ropes observed at 1 AU to the amount of erosion undergone by the MCs in interplanetary space.}
 {The twist profiles of two well-identified MC flux ropes associated with the clear appearance of post eruption arcades in the solar corona are analysed. To infer the amount of erosion, the magnetic flux content of the ropes in the solar atmosphere is estimated, and compared to estimates at 1 AU.}
  {The first MC shows a monotonically decreasing twist from the axis to periphery, while the second displays high twist at the axis, rising twist near the edges, and lower twist in between. The first MC displays a larger reduction in magnetic flux between the Sun and 1 AU, suggesting more erosion, than that seen in the second MC.}
  {In the second cloud, rising twist at the rope edges may have been due to an envelope of overlying coronal field lines with relatively high twist, formed by reconnection beneath the erupting flux rope in the low corona. This high-twist envelope remained almost intact from the Sun to 1 AU due to the low erosion levels. In contrast, the high-twist envelope of the first cloud may have been entirely peeled away via erosion by the time it reaches 1 AU.}


   \maketitle
%

\section{Introduction}
Coronal mass ejections (CMEs; \citeauthor{Webb2012}, \citeyear{Webb2012}) are enormous expulsions of plasma and magnetic flux from the Sun into the heliosphere. The basic structure of the magnetic field of a CME as it erupts is that of a large-scale magnetic flux rope (FR). In interplanetary space, CME-associated FRs that have enhanced magnetic field intensities, smoothly rotating magnetic field vectors, and low proton temperature \citep{1981JGR....86.6673B,1982JGR....87..613K,burlaga1995interplanetary} are called magnetic clouds (MCs). Due to their strong magnetic field intensities, high speed and potential for supporting sustained southward magnetic fields, MCs drive the most intense geomagnetic storms \citep[e.g.][]{kilpua2017geoeffective}. The coherent rotation of magnetic field vector observed inside MCs as it passes the spacecraft represents the systematic twist of the field lines as they wind around the FR central axis. Twist is an intrinsic property of magnetic flux ropes that is related to the stability of FRs. The distribution of twist has important consequences for energetic particle propagation inside FRs because the twist modifies the length of the FR field lines \citep[e.g.][]{larson1997tracing}. Along with magnetic field intensity, axis orientation, the field line twist -- winding of magnetic field lines around MC axis, and chirality -- the right-handed or left-handed sense of twist of FRs are also important FR properties affecting their geoeffectiveness.
\par
 Twist distribution also has important implications for the formation of FRs. There is a long-standing debate on whether FRs are formed during the eruption of a CME due to magnetic reconnection \citep{antiochos1999model,karpen2012mechanisms,moore2001onset} or whether they are pre-existing in the corona prior to the eruption \citep{kopp1976magnetic,titov1999basic}. FRs formed during eruption might have a twist profile where the twist increases gradually from the axis to the periphery of the FR as illustrated in \citet{moore2001onset}. In the case of FRs existing prior to the eruption, magnetic flux can also be added during the eruption to the 'seed FR' due to reconnection occurring beneath the FR. In this case, the reconnected field lines below the FR that are connected to the solar surface form post-eruption arcades (PEAs) and flare ribbons \citep{priest2017flux}. In contrast, the reconnected field lines above the reconnection site envelop the erupting FR, forming a new outer shell for the FR \citep[e.g.,][]{longcope2007quantitative}. The twist profiles in such FRs  may show different twists at the core and outer shell. The pre-existing core or seed FR is generally assumed to be highly twisted  \citep{priest2017flux}, while some studies find that the field lines added during the eruption have low twist \citep{van1989formation}. Other studies, however, have suggested that the outer envelope should also have high twist \citep{2007ApJ...669..621L}. \citet{0004-637X-851-2-123} suggests that the helicity added to the FR through low-corona reconnection is broadly consistent with the helicity of FRs measured in near-Earth space. While coronal reconnection plays a significant role in shaping the magnetic field of the erupting plasma, the details of the processes are still debated.
One possible FR configuration in its minimum energy state is given by the Lundquist model \citep{lundquist1950magnetohydrostatic,2006AnGeo..24..215L}, a linear force-free cylindrical and axially symmetric set of solutions for the FR magnetic field. The Lundquist model is one of a number that have been proposed to describe the inner magnetic structure of MCs \citep{Kilpua2017ICME}. In the Lundquist model, the twist of the FR increases towards the FR boundaries as axial field intensity decreases. MCs have also been modeled as non-linear force-free FRs with uniform twist throughout using the Gold-Hoyle (GH) model \citep{gold1960origin}. Both Lundquist and GH solutions have been successfully fitted to several magnetic field time series in MCs at different distances from the Sun, and have also both produced  good fits for the same MC \citep{wang2016twists,Good2019,Kilpua2019Frontiers}. Studies of twist profiles in MCs have in contrast shown the conflict between these and other models, since the twist distribution obtained depends on the assumed model. For example, \cite{Mostl09}
studied an MC observed on May 20-21, 2007 using the Grad--Shafranov reconstruction technique and found that the twist first increased from the FR core outwards, but then declined, i.e. opposite to the twist profile in the Lundquist solution. \cite{hu2015magnetic} found similar results in a statistical study that used Grad--Shafranov reconstruction and field line length estimates from solar energetic particle observations. \cite{wang2018understanding} and \cite{zhao2018twist} on the other hand found that twist decreases monotonically from the axis to the periphery of FRs using a velocity-modified GH model \citep{wang2016twists} that considers dynamic evolution of MCs. In contrast, \citet{lanabere2020magnetic} recently applied a superposed epoch analysis to MCs and analyzed their magnetic components in the FR frame, finding that the typical twist distribution of MCs is nearly uniform across the central region and increases moderately (by up to a factor of two) towards MC periphery.
\par
Regardless of their intrinsic configuration, the magnetic flux and helicity of FRs may be affected by interaction with the ambient solar wind magnetic field as they propagate outwards in interplanetary space. Such interaction may occur via magnetic reconnection when FR fields are oppositely directed to the local heliospheric magnetic field, which can drape around the FR during its interplanetary propagation \citep{1987GeoRL..14..355G,2015JGRA..120...43R,pal2020flux}. Several studies suggest that reconnection may occur both at the front or back of FRs \citep{2015JGRA..120...43R,pal2020flux}. Reconnection can decrease the flux of the FR by peeling off the FR outer layers. It removes magnetic field lines from the FR outer shell and creates a corresponding number of open field lines \citep{dasso2006new}. It is evident from \citet{dasso2006new} that erosion may remove flux and helicity from FRs and result in asymmetry in the FR azimuthal flux. \citet{pal2020flux} further showed that FR erosion is modulated by the solar cycle and may affect the geoeffectiveness of FRs.\par
 In this paper, we investigate the effect of FR erosion on twist, an intrinsic FR property. Using in situ magnetic field and plasma data in a frame of reference determined by the FR axis direction, we derive the distribution of twist in the cross-sections of two FRs. In the Sun-Earth domain, we analyze FR magnetic flux in a plane formed by the FR axis and spacecraft trajectory to determine how much magnetic flux is eroded during interplanetary propagation. Finally, we demonstrate how erosion impacts the FR twist profiles. In Section 2, the events and methodology selected for this study are described. In Section 3, the results of the analysis are presented, with discussion and conclusions in the final sections.

\section{Overview of Events and Methodology}
\subsection{Event selection and observations}
For this study we select two events with clearly identified MC structures at 1 AU and distinct CMEs accompanied by post eruption arcades (PEAs). The selected events are also required to have (1) a small perpendicular distance $d$ between MC axis and spacecraft trajectory (i.e., small impact parameter $p$) at 1 AU so that the spacecraft provides a complete sampling from the periphery to the core of the MC, and (2) unambiguously identified front and rear MC boundaries to give accurate least-squares fitting of the MCs.
The selected MCs were observed on April 5--6, 2010 (Event 1) and on July 13--14, 2013 (Event 2). Event 1 was the first geoeffective event of solar cycle 24 \citep{wood2011empirical} that caused radio bursts, solar energetic particle (SEP) events, and a prolonged geomagnetic storm with minimum $D_{st}$ of -72 nT. It resulted in a breakdown of the $Galaxy 15$ satellite. Several studies \citep{mostl2010stereo, liu2011solar, wood2011empirical} have analysed solar source, Sun-Earth propagation, kinematics and morphology of this event. Event 2 resulted in a moderate geomagnetic storm with minimum $D_{st}$ of -81 nT. \citet{lugaz2020evolution} analysed the coronal and heliospheric observations of this event, and concluded that its long duration ($\approx$39 hours) was due to expansion in the corona and innermost heliosphere, rather than the result of rapid heliospheric expansion between Mercury at 0.45 AU and Earth at 1 AU. The front and rear boundaries of both the MCs considered here are very similar to those identified in previous studies, namely \citet{mostl2010stereo, liu2011solar, kilpua2017coronal, palmerio2018coronal,mostl2018forward, lugaz2020evolution} and available databases such as HELCATS (\url{https://www.helcats-fp7.eu/catalogues/wp4_icmecat.html}), the \citet{Richardson2010} catalogue (\url{http://www.srl.caltech.edu/ACE/ASC/DATA/level3/icmetable2.htm}) and the Wind ICME catalogue (\url{https://wind.nasa.gov/ICME_catalog/ICME_catalog_viewer.php}). We quantify the asymmetry ($C_B$) in the magnetic field profile of the two MCs following a procedure described in \citet{lanabere2020magnetic}. To avoid asymmetric events in the FR twist calculation, the study sets the constraint to $|C_B|\leq 0.1$. In our study, the values of $C_B$ for Event 1 and Event 2 are negative (i.e. stronger magnetic field exists at the MC front) and almost equal ($C_B \approx -0.04$). 
\par
The CMEs associated with Event 1 and 2 occurred on April 3, 2010 and July 9, 2013 for Event 1 and Event 2, respectively. The location of the photopsheric magnetic field region involved in eruptions corresponding to Event 1 and 2 were S25E00 and N19E14, respectively, i.e., both located close to the disk centre of the Sun. The photospheric eruption region associated with Event 1 was classified as a $\beta$ type active region and identified with NOAA number 11059, whereas the  region associated with Event 2 was not identified with a NOAA number.  

\par

Magnetic field and plasma data at 64~s resolution were obtained from the Advanced Composition Explorer (ACE) spacecraft’s Solar wind Electron, Proton and Alpha Monitor (SWEPAM) and Magnetic Field Experiment (MAG) instruments. The MCs were identified using the standard \citep{1981JGR....86.6673B} definition, i.e., throughout the MC interval, solar wind magnetic field intensity $B$ is enhanced with respect to that of the ambient solar wind, a smooth rotation exists in the magnetic field components $B_x$, $B_y$ and $B_z$, the plasma-$\beta$ is less than one, and plasma temperature $T_p$ is less than the expected ambient solar wind temperature $T_{ex}$ \citep[as subsequently derived by][]{lopez1986solar}. The MC start ($t_{MC,s}$) and end times ($t_{MC,e}$) are provided in columns 2 and 3 of Table \ref{t1}, respectively where column 1 indicates the event number.  \par
We observe that the two events analyzed here were associated with distinct halo CMEs detected by the C2 coronagraph of the Large Angle and Spectrometric Coronagraph (LASCO) on board the Solar and Heliospheric Observatory \citep[SOHO][]{domingo1995soho}. The halo CMEs associated with Event 1 and 2 were first detected by the LASCO/C2 coronagraph at 10:33 UT and 15:12 UT on April 3, 2010 and July 9, 2013, respectively. Both CMEs left behind PEAs as coronal signatures. The eruption associated with Event 1 on April 3, 2010 was accompanied by an eruptive filament and a B-class flare, whereas the one related to Event 2 on July 9, 2013 only followed an eruptive filament. The associated PEAs were observed using Extreme Ultra-Violet (EUV) observations from the Atmospheric Imaging Assembly (AIA) and Extreme ultraviolet Imaging Telescope (EIT) onboard the solar Dynamics Observatory (SDO) and SOHO, respectively. The solar source location of the progenitor CMEs were obtained from \url{https://cdaw.gsfc.nasa.gov/CME_list/halo/halo.html}.

\subsection{Analysis Methods}
In this section, the methods used to determine the twist and magnetic flux in the FRs observed in situ at 1~AU and near the Sun are explained. In addition, the method for determining magnetic flux loss during propagation from the Sun to 1~AU is described.

\subsubsection{Determination of MC twist profile}

\begin{figure}[htbp]
 \centering
 \includegraphics[width=.5\textwidth]{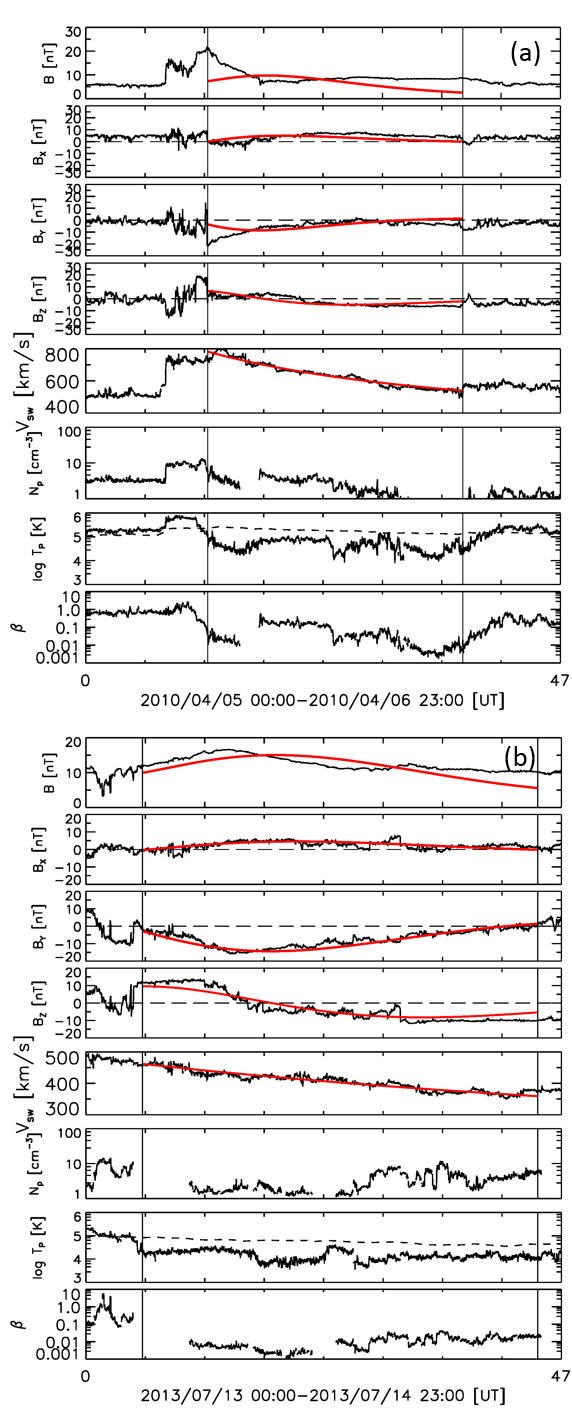}
 \caption{Observations of MCs associated with (a) Event 1 and (b) Event 2 measured by the ACE spacecraft. From top to bottom, the panels show the magnetic field intensity $B$, the three magnetic field components $B_x$, $B_y$, and $B_z$ in GSE coordinates, plasma velocity $V_{sw}$, proton density $N_p$, temperature $T_p$, and plasma-$\beta$. The dashed curves plotted over $T_p$ represent $T_{ex}$. The vertical black lines indicate the start and end times of the MCs. Time is given in hour units on the x axes; the times in UT corresponding to the zero-hour times are indicated below the axes.}
 \label{f1}
\end{figure}

The MCs are assumed to be cylindrical FRs and their twists, $\tau$, as a function of radius, $r$, are calculated following \cite{lanabere2020magnetic}:
\begin{equation}
\tau(r)=B_{\theta,FR}/rB_{z,FR},
    \label{eq1}
\end{equation}
where $B_{\theta,FR}$ and $B_{z,FR}$ are the azimuthal and axial magnetic field components, respectively. The radial distance from the FR axis is given by $r = \sqrt{x^2_{t}+(pr_0)^2}$, where the impact parameter, $p$, is defined as the perpendicular distance between the FR axis and spacecraft trajectory normalised to the FR radius $r_0$, and $x_t$ is the distance travelled by the spacecraft through the MC during time $t$ in a plane perpendicular to its axis. In order to compute $\tau(r)$ from the axial and azimuthal field components, the Cartesian components $B_x$, $B_y$ and $B_z$ in geocentric solar ecliptic (GSE) coordinates are converted to the FR frame attached to the MC axis, with axial component $B_{z,FR}$ and orthogonal components $B_{x,FR}$ and $B_{y,FR}$. As $B_{x,FR}$ and $B_{y,FR}$ are projections of $B_{\theta,FR}$, Equation \ref{eq1} can be written as 
\begin{equation}
\tau(r)=\frac{\sqrt{B^2_{x,FR}+B^2_{y,FR}}}{rB_{z,FR}}.
    \label{eq2}
\end{equation}
To compute the twist profile, $x_t$ is initialized to zero at the FR axis, where $B_{y,FR}$ changes its sign. Equation \ref{eq2} suggests an enhancement in $\tau$ near the axis where $x_t$ approaches zero and impact parameter $p\approx 0$. To determine $B_{x,FR}$, $B_{y,FR}$ and $B_{z,FR}$, knowledge of the FR axis orientation in terms of latitude $\theta_{FR}$ and longitude $\phi_{FR}$ is required, where $\theta_{FR}$ is the angle between the FR axis and ecliptic plane and $\phi_{FR}$ is the angle between the FR axis projected onto the ecliptic plane and the Sun-Earth line. To determine $\theta_{FR}$, $\phi_{FR}$, $r_0$ and $p$, we apply a least-squares FR fit (FRF) to the in situ measurements using a linear force-free model that solves \citep{1988JGR....93.7217B, 1986AdSpR...6..335M} $\nabla \times \mathbf{B}=\alpha \mathbf{B}$ in a cylindrical coordinate system \citep{1990JGR....9511957L} during MC intervals and involves FR expansions \citep{2007AnGeo..25.2453M}. Here $\mathbf{B}$ is the magnetic field vector, and $\alpha$ is a constant that allows self-similar expansion to a cylindrical FR structure. Once $\theta_{FR}$ and $\phi_{FR}$ are derived, the field components in GSE coordinates are rotated to FR frame ($\hat{x}_{FR}$, $\hat{y}_{FR}$, $\hat{z}_{FR}$). In the FR frame, $\hat{z}_{FR}$ is along FR axis with $B_{z,FR}>0$ at the axis, $\hat{y}_{FR}$ is towards the direction of $\hat{z}_{FR} \times \hat{d}$, where $\hat{d}$ is the direction of the spacecraft's rectilinear trajectory, and $\hat{x}_{FR}$ completes the right-handed frame. In Figure \ref{f1}(a) and (b), the force-free fit to the selected MCs are shown with red curves over-plotted on the observed profiles, shown in black. The magnetic field components $B_x$, $B_y$ and $B_z$ are plotted in GSE coordinates. The root mean square error $E_{rms}$ between the observed and modelled FR magnetic field profiles are $E_{rms}$ = 0.32 for Event 1 and 0.24 for Event 2. Figure \ref{f1}(a) and (b) show the solar wind plasma and magnetic field parameters corresponding to Event 1 and Event 2, respectively.

\subsubsection{Determination of FR azimuthal flux}

When FRs form and erupt from the Sun, they accumulate a certain amount of magnetic flux in the azimuthal plane. We refer here to this `initial' azimuthal flux close to the Sun as $\phi_{az,FR}$. During interplanetary propagation, FRs can interact and reconnect with the ambient interplanetary magnetic field. Reconnection of FR field lines can lead to substantial erosion of FR flux in the azimuthal plane (i.e., the plane formed by the FR axis and spacecraft trajectory in the FR frame), and therefore, to the imbalance of the azimuthal flux \citep{2012JGRA..117.9101R,2015JGRA..120...43R,pal2020flux}. In contrast, if no erosion occurs during interplanetary propagation and flux is conserved in the FR azimuthal plane, the MC azimuthal flux measured at 1 AU in situ ideally should be equal to the initial FR flux. Erosion of FRs is expected to also influence the twist profile of the FRs. The amount of reconnected flux and the flux imbalance that results from erosion can be estimated from in-situ observations. However, the imbalance estimated from in-situ observations does not  necessarily give the total reconnected flux due to interplanetary propagation, since the reconnected field lines may completely detach from FRs and create a similar amount of open field lines. Therefore, to estimate how much the FR has been eroded during its whole Sun-to-Earth transit, fluxes near the Sun ($\phi_{az,FR}$) and in-situ at 1 AU ($\phi_{az,MC}$) must be compared. We here employ a technique called the `direct method' \citep{dasso2006new, 2007SoPh..244..115D} to estimate the magnetic flux $\phi_{az,MC}$ (per unit length) that an eroded FR contains upon reaching 1 AU. For this method to be valid, it is necessary that the reconnected field lines are still attached to the FR. In the direct method, the accumulated azimuthal magnetic flux per unit length $\phi_{y,MC}(t_1,t_2)/L$ can be estimated using 
\begin{equation}
    \frac{\phi_{y}(t_1,t_2)}{L}=\int_{t_1}^{t_2}B_{y,FR}(t)V_{x,FR}dt,
   \label{eq3}
\end{equation}
where $V_{x,FR}$ represents the FR speed in the direction of $\hat{x}_{FR}$ and $L$ is the length of the FR. To compute $\phi_{az,MC}$ of an eroded FR at 1 AU, $t_1$ and $t_2$ of Equation \ref{eq3} are considered as $t_{front}$ ($t_{rear}$) and $t_{centre}$ -- a time when $B_{y,FR}$ changes its sign if the FR's reconnected field lines are accumulated at its rear (front). The error in flux determination results from ambiguity in the FR boundaries that can impact the model used to determine FR axis orientation \citep{2003JGRA..108.1356L}.    \par
The initial azimuthal flux $\phi_{az,FR}$ can be alternatively estimated using remote-sensing solar images. Here, we compute the low-coronal magnetic reconnection flux $\phi_{rec}$. Several studies demonstrate that PEAs formed during eruption map to low-coronal reconnection regions associated with eruptive FRs \citep{2007ApJ...669..621L,2007ApJ...659..758Q,2014ApJ...793...53H}. \citet{longcope2007modeling} show that $\phi_{rec}$ approximately equals $\phi_{az,FR}$. We derive $\phi_{rec}$ following the procedure developed by \citep{2017SoPh..292...65G}. Using EUV images, a full-grown PEA at the eruption location on the solar disc is first identified. PEA footpoints are indicated on EUV images, and the footpoints are overlaid on the associated magnetograms. Then, magnetic flux is computed using the magnetic field intensity and the area of the region surrounded by the overlaid PEA footpoints. Here,  $\phi_{rec}$ is measured following

\begin{equation}
    \phi_{rec}=\frac{1}{2}\int_{PEA} \mid B_{los}\mid da,
   \label{eq4}
\end{equation}
where $B_{los}$ represents the SDO/HMI line-of-sight (LOS) magnetic field component corresponding to the region surrounded by PEA footpoints and $da$ is the elemental PEA area. In this study, the PEA structures are identified using SOHO/EIT 195 \AA\ and SDO/AIA 193 \AA \ images. To determine the 1-$\sigma$ error $\delta\phi_{rec}$ involved in estimating $\phi_{rec}$, we have selected PEAs for multiple times during the interval when full-grown PEAs appear in the solar EUV images.

\section{Analysis and Results}
We have computed the azimuthal fluxes for the Event 1 and 2 FRs in the near-Sun region (initial flux) and at 1 AU using Equations \ref{eq4} and \ref{eq3}, respectively. The azimuthal flux $\phi_{az,MC}$ of both the events is calculated using the axis orientations estimated by FRF. By comparing the fluxes of FRs in these two domains, any erosion that occurred during interplanetary propagation can be determined. In Column 9 of Table \ref{t1} we present the azimuthal flux of FRs estimated at 1 AU. Twist profiles have been derived with Equation \ref{eq2}, and the effect of erosion on the field line twists has been analysed. \par

\begin{figure}[htbp]
 \centering
 \includegraphics[width=1.0\textwidth]{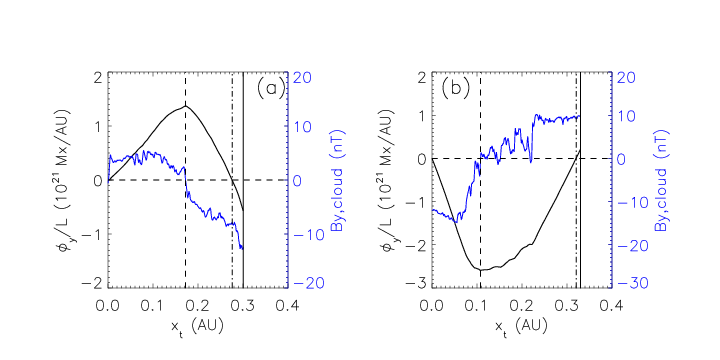}
 \caption{Accumulated azimuthal flux $\phi_{y}/L$ (black curve) and azimuthal magnetic field component (blue curve) plotted with respect to the distance $x_t$ associated with (a) Event 1 and (b) Event 2. Vertical black dashed lines and black dash-dotted lines correspond to $t_{center}$ and $x_{asym}$, respectively. The vertical solid lines represent the MC boundaries.   }
 \label{f2}
\end{figure}

In Figure \ref{f2}(a) and (b), we show the accumulated azimuthal flux per unit length $\phi_{y}/L$ (black curve) and azimuthal magnetic field component $B_{y,FR}$ (blue curve) corresponding to Event 1 and Event 2, respectively, where the field and velocity components in the FR frame are estimated using FR orientation derived from the FRF method. The x-axis of the plots represents the distance $x_t$. The $x_t$ is initialized to zero at the MC front boundary. The imbalance in $\phi_{y}/L$ indicates that both MCs underwent erosion while crossing the spacecraft. The vertical black dashed lines over-plotted on Figure \ref{f2}(a) and (b) indicate the peak of $\phi_{y}/L$, corresponding to the time $t_{center}$ when the spacecraft crossed almost through the center of the MCs in the FR frame. The vertical dash-dotted black lines over-plotted on Figure \ref{f2}(a) and (b) indicate the radial distances $x_{asym}$ when the asymmetry in $\phi_{y}/L$ curves begin. Note that if $p$ is too small, $x_{asym}$ is almost equal to the radius of the eroded MCs in which reconnected field lines are excluded. With the solid vertical lines the boundary of FRs are indicated.

\par
\begin{figure}[htbp]
 \centering
 \includegraphics[width=1.0\textwidth]{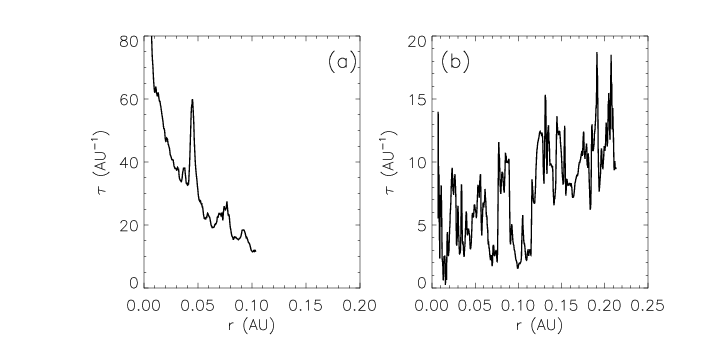}
 \caption{Twist ($\tau$) profile with respect to the radial distance $r$ of MCs associated with (a) Event 1 and (b) Event 2. The plots start from MC centres.} 
 \label{f3}
\end{figure}

The twist $\tau$ of Events 1 and 2 as a function of the radial distance $r$, initialized to zero at the FR centre, are presented in Figure \ref{f3}(a) and (b), respectively. To avoid significant fluctuations introduced by small scale irregularities in FR magnetic field and local distortions in the FR twist profile, we apply a 10-minute forward moving average to the time series of the FR magnetic field components. Similar to \citet{wang2016twists, wang2018understanding, zhao2018twist}, both events have twist profiles that show an increase towards the FR axis. We observe that the twist in the cross-section of Event 1 monotonically decreases with increasing $r$ and acquires a value of $\tau \approx 12$ AU$^{-1}$ at its periphery. The twist profile of Event 2 decreases from the FR centre and fluctuates with mean $\tau \approx 5$ AU$^{-1}$ until half of its radius $r \approx 0.11$ AU. Then the twist increases towards the cloud's outer boundary and reaches a value of $\tau \approx 9$ AU$^{-1}$ at its periphery. The high twist at the edges of Event 2 is consistent with the Lundquist FR model, but few studies \citep[e.g.][]{wang2016twists, wang2018understanding} have recently argued that the high twist observed at the centre of FRs is not consistent with the Lundquist flux rope, which has a low-twist core.
 
\par
\begin{figure}[htbp]
 \centering
 \includegraphics[width=0.7\textwidth]{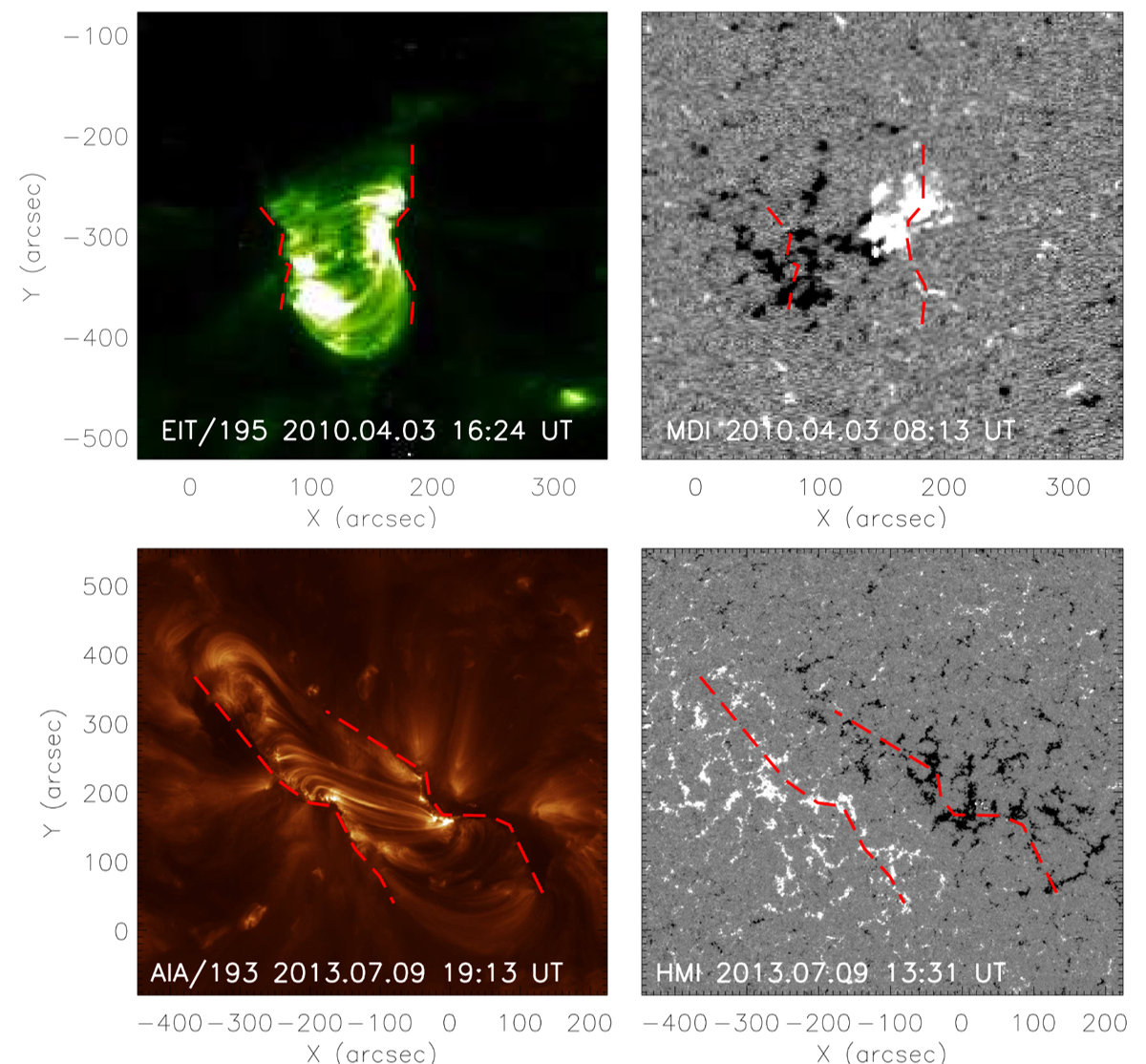}
 \caption{Post-eruption arcades (PEAs) observed in EUV wavelengths at the solar sources of FRs corresponding to Event 1 (upper panel) and 2 (lower panel). Arcade footpoints are marked using red dashed lines on both EUV images and corresponding LOS magnetograms. The PEA shown in the upper panel is associated with a filament and a flare eruption whereas the PEA in the lower panel is associated with a filament eruption only. No flare event was observed during the eruption of the Event 2 CME.  }
 \label{f4}
\end{figure}

The initial azimuthal flux of FRs $\phi_{az,FR}$ is estimated by measuring $\phi_{rec}$. In the upper and lower panels of Figure \ref{f4}, the EUV and LOS magnetograms of the source region of Event 1 and 2 are shown, respectively. The full-grown PEAs associated with Event 1 (detected at 16:24 UT on April 3, 2010) and 2 (detected at 19:13 UT on July 9, 2013), are observed in SOHO/EIT 195 \AA \ and SDO/AIA 195 \AA \ images, respectively. The PEA footpoints are indicated with red dashed lines on both EUV images and magnetograms. To estimate the error involved in measuring $\phi_{rec}$, the PEAs are observed for an interval during which they appear on solar EUV images almost with their full-grown structures. The interval for Event 1 and 2 are chosen as 15:36 - 18:12 UT on April 3, 2010 and 18:00 - 20:30 UT on July 9, 2013, respectively. The corresponding $\phi_{rec}$ for Event 1 and 2 are thus derived, and then averaged. The average initial azimuthal flux $\bar{\phi}_{az,FR}$ of the FRs are estimated from the average $\phi_{rec}$ and the standard deviation of the $\phi_{rec}$ values are calculated to obtain the 1-$\sigma$ error, $\delta \phi_{az,FR}$. In Column 10 of Table \ref{t1}, $\bar{\phi}_{az,FR}$ and its 1-$\sigma$ error $\delta\phi_{az,FR}$ associated with Event 1 and 2 are provided. The percentage of erosion $Er$ of FRs during Sun-Earth travel is derived using $\frac{\bar{\phi}_{az,FR}-\phi_{az,MC}}{\bar{\phi}_{az,FR}} \times 100$. The $Er$ corresponding to Event 1 and 2 are listed in Column 11 of Table \ref{t1}.

\begin{table*}
\caption{The FR start and end times, orientation, impact parameter and radius estimated using 
FRF methods along with the 1-AU and near-Sun azimuthal magnetic flux of FRs and their percentage of erosion during Sun-Earth propagation.} 
\centering
 \begin{tabular}{ccccccccccc}
 \hline        
Event&$t_{MC_s}$&$t_{MC_e}$&$F_T$&$\theta_{FR}$&$\phi_{FR}$&$p$&$r_0$&$\phi_{az,MC}$&$\bar{\phi}_{az,FR}\pm\delta \phi_{az,FR}$&$Er$\\ 
&(UT)&(UT)&&$(^{\circ})$&$(^{\circ})$&&(AU)&($\times 10^{21}$ Mx)&($\times 10^{21}$ Mx)&(\%)\\ \hline

1 &2010 Apr 05 &2010 Apr 06 &FRF &$-24.7$&305.5&$-0.01$&$0.13$&1.4 &2.9 $\pm$0.1&54 \\
&12:05 &13:20&&&&&&&&\\ 
2&2013 Jul 13 &	2013 Jul 14& FRF&$-14.5$&292&0.04&0.15&2.6&3.3 $\pm$0.2&22\\
&05:35&20:40 &&&&&&&& \\
\hline

\label{t1}
\end{tabular}
\end{table*}

\section{Discussion}
This work investigates the influence of FR erosion on FR twist profiles. For this analysis, we chose two FRs having clear boundaries, a small perpendicular distance between the spacecraft trajectories and FR axes, and well-identified PEAs at their solar sources. We note that the fit for Event 1 does not capture the magnetic field magnitude profile very well (Figure \ref{f1}(a)). However, it is common for fitting models to struggle with reproducing the magnitude although they can find acceptable values for most of the free parameters \citep{lepping2018magnetic}. Furthermore, only the axis orientation from the fit is used, and the subsequent flux calculations were performed with the `direct method’ (Equation 3), which uses the measured field and not the fitted profile. The fit is used to place the data in a coordinate system relative to the axis orientation. The fit captures the axis orientation reasonably well (the polarities of the individual components are accurately reproduced), and the $E_{rms}=0.32$. The fit for Event 2 is very good in terms of the components, and it also reproduces the magnitude profile quite well, with an overall $E_{rms}=0.24$.

\par
Using superposed epoch analysis (SEA), which emphasizes common features in MC profiles, \citet{lanabere2020magnetic} determined a typical twist distribution inside MCs. The distribution is uniform in the centre-half part of FRs and gradually increases by up to a factor of two towards the FR boundaries. The enhancement of twist in the outer shell of FRs may result from helical field lines formed from low-coronal magnetic reconnection beneath erupting FRs at the solar sources that wrap around the FR core \citep{2007ApJ...659..758Q}. However, we note that, as discussed in the Introduction, some studies suggest a lower twist in the envelope and higher twist in the core. In our study, the twist profile of Event 2 increases moderately towards the periphery from a uniform value in the central part. In contrast, the twist of Event 1 does not show any enhancement at the periphery. 
The twist profiles we have derived here mostly cover the inner part of FR core, i.e., the part close to the FR axis. At this inner core, we observe in both events a substantial increase in twist towards the center of the FRs, similar to that found by imposing a velocity-modified, uniform-twist GH model to MCs \citep{wang2018understanding, zhao2018twist}. \par
To account for the dissimilarity in the twist profiles of the two events, we determine their azimuthal magnetic flux both near the Sun and in situ (i.e. $\sim 1$ AU), and in particular consider the effect of erosion due to magnetic reconnection between the FR fields and the ambient field (see the Introduction).  The reconnection rate is expected to be high close to the Sun and in the inner heliosphere \citep{2008JGRA..113.0B08L}. \citet{2014JGRA..119...26L} showed that 47-67\% erosion is expected to occur within $\approx$ 0.39 AU. However, the available in situ observations of FRs do not allow us to locate the reconnection site in the interplanetary medium. One of the signatures of the ongoing erosion of FRs is the imbalance in azimuthal flux captured at the time of in situ observation. An asymmetry in azimuthal flux due to reconnection close to the Sun may not be identified by in situ observations because, at the time of observation at 1 AU, the reconnected field lines may completely detach from the FRs and create a similar number of open field lines. Therefore, we compare near-Sun and near-Earth azimuthal flux of FRs to determine whether FRs are significantly peeled off during their Sun-Earth propagation. We find that the percentage of erosion $Er$ for Event 1 is greater than that of Event 2. Furthermore, \citep{mostl2010stereo} concluded that the long-duration, non-rotating magnetic field observed at the back part of Event 1 FR is very reminiscent of the events studied by \citet{2007SoPh..244..115D} where the long back region with non-rotating magnetic fields behind MCs are resulted from reconnection between MC and the interplanetary magnetic field \citep{2007SoPh..244..115D}.

The dissimilarity in twist profiles observed here may result from the difference in their erosion. We suggest that a significant erosion may have completely removed the twisted outer layer of the FR associated with Event 1. Therefore, for this event, only the outer core along with the inner core remained by the time it reached the orbit of the Earth. As erosion is less in Event 2 than Event 1, the FR's twisted outer shell is only partly removed and it results in a twist profile that rises towards the FR periphery. \citet{wang2018understanding} found a twist profile monotonically decreasing from core to periphery, which supports the scenario that a twisted seed FR exists prior to the eruption. Note that the study found the degree of imbalance of the MC as 25\% and suggested that the erosion affected the MC twist. As the MC was a slow and weak event, its source CME could not be distinguished from other preceding and following CMEs. Its solar source location was thus ambiguous, and therefore, the \citet{wang2018understanding} study cannot confirm whether a twisted MC outer shell existed before its erosion in interplanetary space.  \par
\begin{figure}[htbp]
 \centering
 \includegraphics[width=1.0\textwidth]{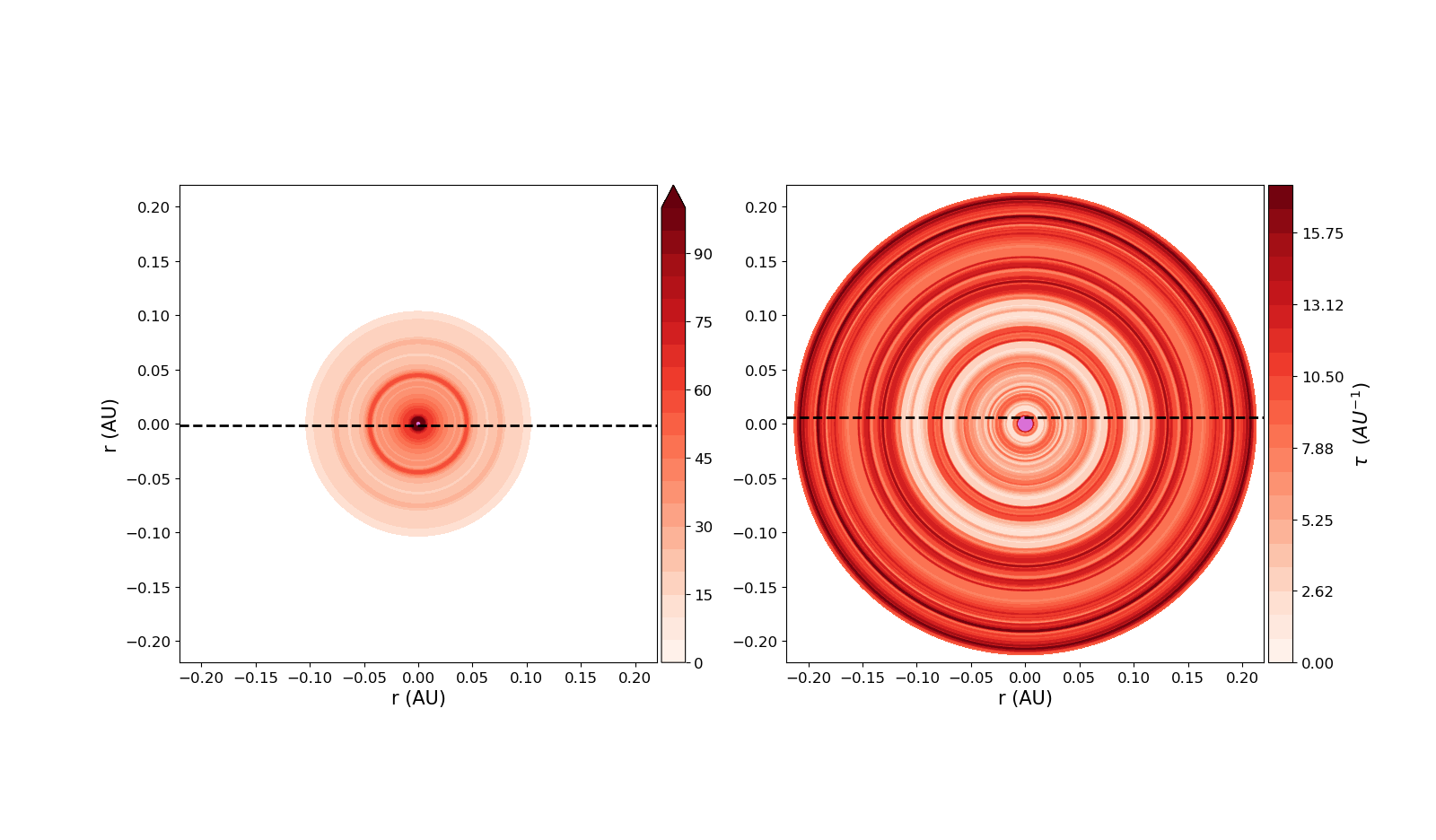}
 \caption{Radial distribution of twist in the cross-section of two MCs replicating Event 1 (left panel) and 2 (right panel) at 1 AU. The color bars represent the twist values and the dashed lines show the spacecraft propagation directions. The light purple color dots at the centre of the cross-sections indicate the portions for which twist values are not available.}
 \label{f5}
\end{figure}

Unlike \citet{zhao2018twist} and \citet{wang2016twists}, the twist of MCs is computed in this study directly from in situ magnetic field components rotated to the FR frame. Here we consider MC azimuthal flux per unit length because a statistical study by \citet{2007ApJ...659..758Q} found that the ratio of MC azimuthal flux to the coronal reconnection flux equals to unity while considering $L=1$ AU. \citet{longcope2007modeling} showed that the total magnetic helicity transported into the FR from coronal sheared arcades due to low coronal reconnection matches well with the magnetic helicity of MCs when the length of the MC is considered as 1 AU.  
In the left and right panels of Figure \ref{f5}, we represent two MC cross-sections with their twist distributions using colored contours. The associated FRs have twist profiles similar to those of Event 1 and 2, and undergo same level of erosion as Event 1 and 2 do while propagating through the interplanetary medium. The highly-twisted outer layer is absent in the FR cross-section of left panel, whereas it remains in the other one. The spacecraft propagation paths are represented by black dashed lines. \par
We note that the twist profiles are partly dependent on the quality of the fits, which provide the FR axis orientation, impact parameter and width. While the fits for two events analysed here are broadly satisfactory (e.g., in terms of $E_{rms}$), more events will need to be analysed in order to draw more firm and general conclusions. In particular, more events with similar twist profiles to Event 1 (which are less commonly observed) would be worthwhile analysing, in order to build a more statistical picture.

\section{Conclusion}
The twist of the magnetic field lines inside FRs has important implications for the geoeffectiveness of FRs. Moreover, the radial distribution of twist provides significant information on FR formation processes. When FRs propagate through the interplanetary medium, their outer shell might be partially or entirely eroded due to interaction with the ambient solar wind magnetic field. The erosion of FRs can significantly alter their magnetic properties, including magnetic flux, twist, and helicity. In this study, we select two distinct FRs having typical plasma and magnetic characteristics with well-defined boundaries and investigate the effect of FR erosion on their magnetic flux and twist. The post eruption arcades formed at the FR solar sources during eruption confirm the presence of envelopes of overlying coronal field lines around the erupting, CME-associated FRs. The envelope includes the FR outer shell, where the field line twist depends on the twist of the overlying coronal field. In our study, the field line twist for Event 2 is nearly uniform across the FR mid-region i.e., outer core and increases in the FR's envelope i.e., outer shell. In contrast, the field line twist of Event 1 shows a decreasing profile near the FR boundary. Both the twist profiles monotonically increases towards the FR axis. The presence of highly twisted field lines around the FR central axis is consistent with twisted seed FRs pre-existing the eruptions. After comparing the FR azimuthal magnetic flux in the Sun-Earth domain, it is inferred that the percentage of erosion is higher for Event 1 ($Er= 54\%$) than Event 2 ($Er= 22\%$), and that the erosion probably removes the highly twisted outer shell of Event 1. Our result demonstrates that FR eruption involving coronal magnetic reconnection at the solar sources forms an envelope around the FR. The field line twists are greater in the FR envelope (outer shell) than that of the FR outer core, where the twist is almost uniform. During interplanetary propagation, magnetic reconnection that causes erosion of FRs can remove altogether the twisted outer shell and leave the FR with its core having decreasing field line twist values towards the periphery. \par

\begin{acknowledgements}
S.P, E.K., and D.P. acknowledge the European Research Council (ERC) under the European Union's Horizon 2020 Research and Innovation Program Project SolMAG 724391. The results presented here have been achieved under the framework of the Finnish Centre of Excellence in Research of Sustainable Space (FORESAIL; Academy of Finland grant numbers 312390), which we gratefully acknowledge. E.K and S.G acknowledge Academy of Finland Project 310445 (SMASH). S.P thanks Dr. Katsuhide Marubashi for providing us with the linear force-free cylindrical model. We thank referee for helpful comments and acknowledge the use of data from SDO, SOHO and ACE.
\end{acknowledgements}

\begin{thebibliography}{}

\bibitem[{Antiochos {et~al.}(1999)Antiochos, DeVore, \&
  Klimchuk}]{antiochos1999model}
Antiochos, S., DeVore, C., \& Klimchuk, J. 1999, The Astrophysical Journal,
  510, 485

\bibitem[{{Burlaga} {et~al.}(1981){Burlaga}, {Sittler}, {Mariani}, \&
  {Schwenn}}]{1981JGR....86.6673B}
{Burlaga}, L., {Sittler}, E., {Mariani}, F., \& {Schwenn}, R. 1981, J.
  Geophys.Res. Space Physics, 86, 6673

\bibitem[{{Burlaga}(1988)}]{1988JGR....93.7217B}
{Burlaga}, L.~F. 1988, J. Geophys.Res. Space Physics, 93, 7217

\bibitem[{Burlaga \& Burlaga(1995)}]{burlaga1995interplanetary}
Burlaga, L.~F. \& Burlaga, L. 1995, Interplanetary magnetohydrodynamics (Oxford
  University Press on Demand)

\bibitem[{Dasso {et~al.}(2006)Dasso, Mandrini, D{\'e}moulin, \&
  Luoni}]{dasso2006new}
Dasso, S., Mandrini, C.~H., D{\'e}moulin, P., \& Luoni, M.~L. 2006, Astronomy
  \& Astrophysics, 455, 349

\bibitem[{{Dasso} {et~al.}(2007){Dasso}, {Nakwacki}, {D{\'e}moulin}, \&
  {Mandrini}}]{2007SoPh..244..115D}
{Dasso}, S., {Nakwacki}, M.~S., {D{\'e}moulin}, P., \& {Mandrini}, C.~H. 2007,
  Sol. Phys., 244, 115

\bibitem[{Domingo {et~al.}(1995)Domingo, Fleck, \& Poland}]{domingo1995soho}
Domingo, V., Fleck, B., \& Poland, A.~I. 1995, Solar Physics, 162, 1

\bibitem[{Gold \& Hoyle(1960)}]{gold1960origin}
Gold, T. \& Hoyle, F. 1960, Monthly Notices of the Royal Astronomical Society,
  120, 89

\bibitem[{{Good} {et~al.}(2019){Good}, {Kilpua}, {LaMoury}, {Forsyth},
  {Eastwood}, \& {M{\"o}stl}}]{Good2019}
{Good}, S.~W., {Kilpua}, E.~K.~J., {LaMoury}, A.~T., {et~al.} 2019, Journal of
  Geophysical Research (Space Physics), 124, 4960

\bibitem[{{Gopalswamy} {et~al.}(2017){Gopalswamy}, {Yashiro}, {Akiyama}, \&
  {Xie}}]{2017SoPh..292...65G}
{Gopalswamy}, N., {Yashiro}, S., {Akiyama}, S., \& {Xie}, H. 2017, \solphys,
  292, 65

\bibitem[{{Gosling} \& {McComas}(1987)}]{1987GeoRL..14..355G}
{Gosling}, J.~T. \& {McComas}, D.~J. 1987, Geophys. Res. Lett., 14, 355

\bibitem[{{Hu} {et~al.}(2014){Hu}, {Qiu}, {Dasgupta}, {Khare}, \&
  {Webb}}]{2014ApJ...793...53H}
{Hu}, Q., {Qiu}, J., {Dasgupta}, B., {Khare}, A., \& {Webb}, G.~M. 2014, \apj,
  793, 53

\bibitem[{Hu {et~al.}(2015)Hu, Qiu, \& Krucker}]{hu2015magnetic}
Hu, Q., Qiu, J., \& Krucker, S. 2015, Journal of Geophysical Research: Space
  Physics, 120, 5266

\bibitem[{Karpen {et~al.}(2012)Karpen, Antiochos, \&
  DeVore}]{karpen2012mechanisms}
Karpen, J.~T., Antiochos, S.~K., \& DeVore, C.~R. 2012, The Astrophysical
  Journal, 760, 81

\bibitem[{Kilpua {et~al.}(2017{\natexlab{a}})Kilpua, Balogh, von Steiger, \&
  Liu}]{kilpua2017geoeffective}
Kilpua, E., Balogh, A., von Steiger, R., \& Liu, Y. 2017{\natexlab{a}}, Space
  Science Reviews, 212, 1271

\bibitem[{Kilpua {et~al.}(2017{\natexlab{b}})Kilpua, Koskinen, \&
  Pulkkinen}]{kilpua2017coronal}
Kilpua, E., Koskinen, H.~E., \& Pulkkinen, T.~I. 2017{\natexlab{b}}, Living
  Reviews in Solar Physics, 14, 1

\bibitem[{{Kilpua} {et~al.}(2017){Kilpua}, {Koskinen}, \&
  {Pulkkinen}}]{Kilpua2017ICME}
{Kilpua}, E., {Koskinen}, H. E.~J., \& {Pulkkinen}, T.~I. 2017, Living Reviews
  in Solar Physics, 14, 5

\bibitem[{{Kilpua} {et~al.}(2019){Kilpua}, {Good}, {Palmerio}, {Asvestari},
  {Lumme}, {Ala-Lahti}, {Kalliokoski}, {Morosan}, {Pomoell}, {Price},
  {Magdaleni{\'c}}, {Poedts}, \& {Futaana}}]{Kilpua2019Frontiers}
{Kilpua}, E. K.~J., {Good}, S.~W., {Palmerio}, E., {et~al.} 2019, Frontiers in
  Astronomy and Space Sciences, 6, 50

\bibitem[{{Klein} \& {Burlaga}(1982)}]{1982JGR....87..613K}
{Klein}, L.~W. \& {Burlaga}, L.~F. 1982, J. Geophys.Res. Space Physics, 87, 613

\bibitem[{Kopp \& Pneuman(1976)}]{kopp1976magnetic}
Kopp, R. \& Pneuman, G. 1976, Solar Physics, 50, 85

\bibitem[{Lanabere {et~al.}(2020)Lanabere, Dasso, D{\'e}moulin, Janvier,
  Rodriguez, \& Mas{\'\i}as-Meza}]{lanabere2020magnetic}
Lanabere, V., Dasso, S., D{\'e}moulin, P., {et~al.} 2020, Astronomy \&
  Astrophysics, 635, A85

\bibitem[{Larson {et~al.}(1997)Larson, Lin, McTiernan, McFadden, Ergun,
  McCarthy, Reme, Sanderson, Kaiser, Lepping, {et~al.}}]{larson1997tracing}
Larson, D., Lin, R., McTiernan, J., {et~al.} 1997, Geophysical research
  letters, 24, 1911

\bibitem[{{Lavraud} \& {Borovsky}(2008)}]{2008JGRA..113.0B08L}
{Lavraud}, B. \& {Borovsky}, J.~E. 2008, J. Geophys.Res. Space Physics, 113,
  A00B08

\bibitem[{{Lavraud} {et~al.}(2014){Lavraud}, {Ruffenach}, {Rouillard},
  {Kajdic}, {Manchester}, \& {Lugaz}}]{2014JGRA..119...26L}
{Lavraud}, B., {Ruffenach}, A., {Rouillard}, A.~P., {et~al.} 2014, J.
  Geophys.Res. Space Physics, 119, 26

\bibitem[{Lepping {et~al.}(2018)Lepping, Wu, Berdichevsky, \&
  Kay}]{lepping2018magnetic}
Lepping, R., Wu, C.-C., Berdichevsky, D., \& Kay, C. 2018, Solar Physics, 293,
  1

\bibitem[{{Lepping} {et~al.}(2003){Lepping}, {Berdichevsky}, \&
  {Ferguson}}]{2003JGRA..108.1356L}
{Lepping}, R.~P., {Berdichevsky}, D.~B., \& {Ferguson}, T.~J. 2003, J.
  Geophys.Res. Space Physics, 108, 1356

\bibitem[{{Lepping} {et~al.}(2006){Lepping}, {Berdichevsky}, {Wu}, {Szabo},
  {Narock}, {Mariani}, {Lazarus}, \& {Quivers}}]{2006AnGeo..24..215L}
{Lepping}, R.~P., {Berdichevsky}, D.~B., {Wu}, C.~C., {et~al.} 2006, Annales
  Geophysicae, 24, 215

\bibitem[{{Lepping} {et~al.}(1990){Lepping}, {Jones}, \&
  {Burlaga}}]{1990JGR....9511957L}
{Lepping}, R.~P., {Jones}, J.~A., \& {Burlaga}, L.~F. 1990, J. Geophys.Res.
  Space Physics, 95, 11957

\bibitem[{Liu {et~al.}(2011)Liu, Luhmann, Bale, \& Lin}]{liu2011solar}
Liu, Y., Luhmann, J.~G., Bale, S.~D., \& Lin, R.~P. 2011, The Astrophysical
  Journal, 734, 84

\bibitem[{Longcope \& Beveridge(2007)}]{longcope2007quantitative}
Longcope, D. \& Beveridge, C. 2007, The Astrophysical Journal, 669, 621

\bibitem[{Longcope {et~al.}(2007)Longcope, Beveridge, Qiu, Ravindra, Barnes, \&
  Dasso}]{longcope2007modeling}
Longcope, D., Beveridge, C., Qiu, J., {et~al.} 2007, Solar Physics, 244, 45

\bibitem[{{Longcope} \& {Beveridge}(2007)}]{2007ApJ...669..621L}
{Longcope}, D.~W. \& {Beveridge}, C. 2007, \apj, 669, 621

\bibitem[{Lopez \& Freeman(1986)}]{lopez1986solar}
Lopez, R.~E. \& Freeman, J.~W. 1986, Journal of Geophysical Research: Space
  Physics, 91, 1701

\bibitem[{Lugaz {et~al.}(2020)Lugaz, Winslow, \& Farrugia}]{lugaz2020evolution}
Lugaz, N., Winslow, R., \& Farrugia, C. 2020, Journal of Geophysical Research:
  Space Physics, 125, e2019JA027213

\bibitem[{Lundquist(1950)}]{lundquist1950magnetohydrostatic}
Lundquist, S. 1950, Ark. Fys., 2, 361

\bibitem[{{Marubashi}(1986)}]{1986AdSpR...6..335M}
{Marubashi}, K. 1986, Advances in Space Research, 6, 335

\bibitem[{{Marubashi} \& {Lepping}(2007)}]{2007AnGeo..25.2453M}
{Marubashi}, K. \& {Lepping}, R.~P. 2007, Annales Geophysicae, 25, 2453

\bibitem[{Moore {et~al.}(2001)Moore, Sterling, Hudson, \&
  Lemen}]{moore2001onset}
Moore, R.~L., Sterling, A.~C., Hudson, H.~S., \& Lemen, J.~R. 2001, The
  Astrophysical Journal, 552, 833

\bibitem[{M{\"o}stl {et~al.}(2018)M{\"o}stl, Amerstorfer, Palmerio, Isavnin,
  Farrugia, Lowder, Winslow, Donnerer, Kilpua, \& Boakes}]{mostl2018forward}
M{\"o}stl, C., Amerstorfer, T., Palmerio, E., {et~al.} 2018, Space Weather, 16,
  216

\bibitem[{{M{\"o}stl} {et~al.}(2009){M{\"o}stl}, {Farrugia}, {Miklenic},
  {Temmer}, {Galvin}, {Luhmann}, {Kilpua}, {Leitner}, {Nieves-Chinchilla},
  {Veronig}, \& {Biernat}}]{Mostl09}
{M{\"o}stl}, C., {Farrugia}, C.~J., {Miklenic}, C., {et~al.} 2009, J. Geophys.
  Res., 114, A04102

\bibitem[{M{\"o}stl {et~al.}(2010)M{\"o}stl, Temmer, Rollett, Farrugia, Liu,
  Veronig, Leitner, Galvin, \& Biernat}]{mostl2010stereo}
M{\"o}stl, C., Temmer, M., Rollett, T., {et~al.} 2010, Geophysical Research
  Letters, 37

\bibitem[{Pal {et~al.}(2020)Pal, Dash, \& Nandy}]{pal2020flux}
Pal, S., Dash, S., \& Nandy, D. 2020, Geophysical Research Letters, 47,
  e2019GL086372

\bibitem[{Pal {et~al.}(2017)Pal, Gopalswamy, Nandy, Akiyama, Yashiro, Makela,
  \& Xie}]{0004-637X-851-2-123}
Pal, S., Gopalswamy, N., Nandy, D., {et~al.} 2017, The Astrophysical Journal,
  851, 123

\bibitem[{Palmerio {et~al.}(2018)Palmerio, Kilpua, M{\"o}stl, Bothmer, James,
  Green, Isavnin, Davies, \& Harrison}]{palmerio2018coronal}
Palmerio, E., Kilpua, E.~K., M{\"o}stl, C., {et~al.} 2018, Space Weather, 16,
  442

\bibitem[{Priest \& Longcope(2017)}]{priest2017flux}
Priest, E.~R. \& Longcope, D. 2017, Solar Physics, 292, 1

\bibitem[{{Qiu} {et~al.}(2007){Qiu}, {Hu}, {Howard}, \&
  {Yurchyshyn}}]{2007ApJ...659..758Q}
{Qiu}, J., {Hu}, Q., {Howard}, T.~A., \& {Yurchyshyn}, V.~B. 2007, \apj, 659,
  758

\bibitem[{Richardson \& Cane(2010)}]{Richardson2010}
Richardson, I.~G. \& Cane, H.~V. 2010, Sol. Phys., 264, 189

\bibitem[{{Ruffenach} {et~al.}(2015){Ruffenach}, {Lavraud}, {Farrugia},
  {D{\'e}moulin}, {Dasso}, {Owens}, {Sauvaud}, {Rouillard}, {Lynnyk},
  {Foullon}, {Savani}, {Luhmann}, \& {Galvin}}]{2015JGRA..120...43R}
{Ruffenach}, A., {Lavraud}, B., {Farrugia}, C.~J., {et~al.} 2015, J.
  Geophys.Res. Space Physics, 120, 43

\bibitem[{{Ruffenach} {et~al.}(2012){Ruffenach}, {Lavraud}, {Owens}, {Sauvaud},
  {Savani}, {Rouillard}, {D{\'e}moulin}, {Foullon}, {Opitz}, {Fedorov},
  {Jacquey}, {G{\'e}not}, {Louarn}, {Luhmann}, {Russell}, {Farrugia}, \&
  {Galvin}}]{2012JGRA..117.9101R}
{Ruffenach}, A., {Lavraud}, B., {Owens}, M.~J., {et~al.} 2012, J. Geophys.Res.
  Space Physics, 117, A09101

\bibitem[{Titov \& D{\'e}moulin(1999)}]{titov1999basic}
Titov, V. \& D{\'e}moulin, P. 1999, Astronomy and Astrophysics, 351, 707

\bibitem[{van Ballegooijen \& Martens(1989)}]{van1989formation}
van Ballegooijen, A.~A. \& Martens, P. 1989, The Astrophysical Journal, 343,
  971

\bibitem[{Wang {et~al.}(2018)Wang, Shen, Liu, Liu, Guo, Li, Xu, Hu, \&
  Zhang}]{wang2018understanding}
Wang, Y., Shen, C., Liu, R., {et~al.} 2018, Journal of Geophysical Research:
  Space Physics, 123, 3238

\bibitem[{Wang {et~al.}(2016)Wang, Zhuang, Hu, Liu, Shen, \&
  Chi}]{wang2016twists}
Wang, Y., Zhuang, B., Hu, Q., {et~al.} 2016, Journal of Geophysical Research:
  Space Physics, 121, 9316

\bibitem[{{Webb} \& {Howard}(2012)}]{Webb2012}
{Webb}, D.~F. \& {Howard}, T.~A. 2012, Living Reviews in Solar Physics, 9, 3

\bibitem[{Wood {et~al.}(2011)Wood, Wu, Howard, Socker, \&
  Rouillard}]{wood2011empirical}
Wood, B., Wu, C.-C., Howard, R., Socker, D., \& Rouillard, A. 2011, The
  Astrophysical Journal, 729, 70

\bibitem[{Zhao {et~al.}(2018)Zhao, Wang, Feng, Xu, Zhao, Zhao, \&
  Hu}]{zhao2018twist}
Zhao, A., Wang, Y., Feng, H., {et~al.} 2018, The Astrophysical Journal Letters,
  869, L13

\end{thebibliography}

\end{document}